\documentclass[cernpreprint, atlasdraft=false, USenglish, texmf, orcidlogo]{atlasdoc}
\usepackage{atlaspackage}
\usepackage{atlasbiblatex}

\usepackage{atlasphysics}
\graphicspath{{logos/}{figures/}}

\usepackage{ANA-HDBS-2021-32-PAPER-defs}

\AtlasTitle{Search for a new \Zp gauge boson via the \procWzp~process in $pp$ collisions at $\sqrt{s}=13$~\TeV with the ATLAS detector}

\AtlasAbstract{%
A search for a new \Zp gauge boson predicted by \Lmt models, based on charged-current Drell--Yan production, \procWzp, is presented.
The data sample used corresponds to an integrated luminosity of 140~fb$^{-1}$ of proton--proton collisions at $\sqrt{s} = 13$~\TeV recorded by the ATLAS detector at the Large Hadron Collider.
The search examines a final state of $3\mu$ plus large missing transverse momentum.
Upper limits are set on the \Zp production cross-section times branching ratio in the mass range of 5--81~\GeV.
After combining with the previous \Zp search using the neutral-current Drell--Yan production with a $4\mu$ final state,
the most stringent exclusion limits to date are achieved in the parameter space of the \Zp coupling strength and mass.
}

\AtlasRefCode{HDBS-2021-32}

\PreprintIdNumber{CERN-EP-2024-042}

\AtlasJournalRef{Phys. Rev. D 110 (2024) 072008}
\AtlasDOI{10.1103/PhysRevD.110.072008}

\hypersetup{pdftitle={ATLAS document},pdfauthor={The ATLAS Collaboration}}

\begin{document}

\maketitle

\section{Introduction}
%

Various extensions of the Standard Model (SM) feature an extra $U(1)$ gauge symmetry and predict a new massive gauge boson,
generally referred to as \Zp~\cite{Leike:1998wr,Langacker_2009}.
Typical benchmark models include the Sequential Standard Model~\cite{Altarelli:1989ff},
Grand Unified Theories based on the $E_6$ gauge group~\cite{London:1986dk},
and left~(L)-right~(R) symmetric extensions of the SM based on the $SU(2)_L \otimes SU(2)_R \otimes U(1)_{B-L}$ gauge group~\cite{Pati:1974yy,Mohapatra:1974gc,Senjanovic:1975rk},
where $B-L$ denotes the difference between baryon and lepton numbers.
For \Zp boson that significantly couples to light quarks, the constraint on its mass from LHC is up to 5 TeV~\cite{EXOT-2018-08,CMS-EXO-19-019}.
Alternatively, $U(1)$ gauge symmetries based on the difference between lepton family numbers are less constrained and are anomaly free~\cite{PhysRevD.44.2118}.
The model based on gauging the difference between $\mu$-lepton number and $\tau$-lepton number, \Lmt, is particularly interesting
since it is the least constrained experimentally, because the \Lmt\ \Zp boson only couples to the second and third generations of leptons.
In recent years, this model has attracted interest in both theoretical and experimental communities~\cite{Elahi_2016,Elahi_2017,PhysRevLett.114.151801}
because it could address some of the reported anomalies,
such as the measured muon anomalous magnetic moment
~\cite{aguillard2023measurement,PhysRevD.73.072003}
and lepton flavor anomalies~\cite{Aaij_2013,PhysRevLett.125.011802, Blake_2017,PhysRevD.108.032002,Amhis_2023}.
In addition, these models also provide a viable solution to dark matter and neutrino mass~\cite{Altmannshofer:2016jzy,Ma_2002,Langacker_2009}.

The interaction between the \Zp boson and the second- and third-generation leptons can be described with the following Lagrangian
\begin{equation*}
L_{Z'} = - {1\over 4} F_{\alpha\beta}F^{\alpha\beta} + {1\over 2} m^2_{Z'} Z'^\alpha Z'_\alpha
- g_{Z'} Z'_\alpha(\bar\ell_2\gamma^\alpha\ell_2 + \mu_L\gamma^\alpha\mu_R
-\bar\ell_3\gamma^\alpha\ell_3 - \tau_L\gamma^\alpha\tau_R ),
\end{equation*}
where $F_{\alpha\beta} = \partial_\alpha Z'_\beta - \partial_\beta Z'_\alpha$ is the \Zp field strength tensor;
$\ell_i = (\nu_i, e_i)^T$ ($i=2,3$, denoting the second and the third generation left-handed lepton doublets);
and $g_{Z'}$ (from hereon referred to as \coupzp )
is the coupling constant of the interaction between the \Zp boson and the SM leptons.
The $Z'$-boson mass, \mZp, and $g$ are the free parameters of the model.
In proton--proton~($pp$) collisions at the Large Hadron Collider~(LHC), the \Zp boson could be produced by final state radiation from
$\mu$, $\nu_{\mu}$, $\tau$ and $\nu_{\tau}$ leptons originating from other physics processes.

This letter presents the first search for a \Lmt\ \Zp boson produced from leptons arising from charged-current Drell--Yan (DY) process,
\procWzp~(see \Fig{\ref{fig:feynman}}), giving a final state of 3$\mu$ plus large missing transverse momentum.
This novel search complements previous analyses by the ATLAS and CMS Collaborations using neutral-current DY process with a 4$\mu$ final state, \procZzp, and has
a much higher cross-section by a factor~of~3--6.
The CMS Collaboration searched for the \Zp boson in the mass range of 5--70~\GeV{} using 77.3~\ifb~of $pp$ collision data at $\sqrt{s}=13$~TeV~\cite{CMS-EXO-18-008}.
The ATLAS Collaboration searched for the \Zp boson with a mass up to 81~\GeV using 139~\ifb of data at $\sqrt{s}=13$~TeV~\cite{HDBS-2018-57}.

\begin{figure}[!ht]
\centering
\includegraphics[width=0.45\textwidth]{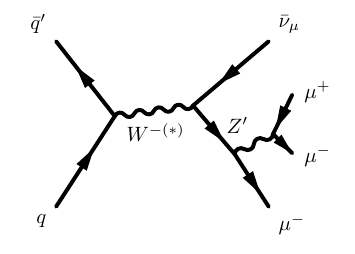}
\label{fig:feynman:a}
\includegraphics[width=0.45\textwidth]{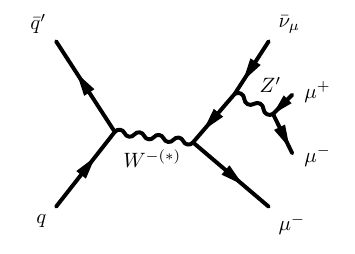}
\label{fig:feynman:b}
\caption{Representative Feynman diagrams of a \Zp boson via radiation off a lepton in charged-current Drell--Yan production giving a $\mu^{-}\mu^{+}\mu^{-}\bar{\nu}_{\mu}$ final state.}
\label{fig:feynman}
\end{figure}


%

%

%
%
%
%
%

%
%
%

%
%

\section{Dataset and Monte Carlo simulation}
%

The ATLAS experiment at the LHC is a multipurpose particle detector
with a forward--backward symmetric cylindrical geometry and a near \(4\pi\) coverage in
solid angle~\cite{PERF-2007-01,ATLAS-TDR-2010-19,PIX-2018-001}.\footnote{ATLAS uses a right-handed coordinate system with its origin at the nominal interaction point (IP) in the center of the detector and the $z$-axis along the beam pipe.The $x$-axis points from the IP to the center of the LHC ring, and the $y$-axis points
upward. Cylindrical coordinates ($r$, $\phi$) are used in the transverse plane, $\phi$ being the azimuthal angle around the $z$-axis.
The pseudorapidity is defined in terms of the polar angle $\theta$ as $\eta = -\ln{\tan{\theta /2}}$.
Angular distance is measured in units of \(\Delta R \equiv \sqrt{(\Delta\eta)^{2} + (\Delta\phi)^{2}}\).}
The $pp$ collision data at $\sqrt{s} = 13$~\TeV\ recorded by the ATLAS experiment during 2015--2018 are used.
The corresponding integrated luminosity is 140~\ifb after applying data quality requirements~\cite{DAPR-2021-01}.
A combination of single-lepton and multi-lepton triggers~\cite{TRIG-2018-01,TRIG-2018-05} is used,
with transverse momentum (\pT) thresholds varying from 20 to 26~\GeV for single-muon triggers, 10 to 14~\GeV for di-muon triggers,
and 6~\GeV for tri-muon triggers. The overall trigger efficiency is greater than 96\% for events passing the offline event selection.
An extensive software suite~\cite{ATL-SOFT-PUB-2021-001} is used in data simulation, in the reconstruction and analysis of real and simulated data, in detector operations, and in the trigger and data acquisition systems of the experiment.

Monte Carlo~(MC)  signal samples are simulated using \MGNLO[2.9.5] \cite{Alwall:2014hca},
with matrix elements~(ME) calculated at leading-order~(LO) in perturbative quantum chromodynamics (QCD) and with the \NNPDF[3.0nlo]~\cite{nnpdf:2014uwa} parton distribution function~(PDF) set.
The events were interfaced to \PYTHIA[8.245]~\cite{Sjostrand:2014zea} to model the parton shower, hadronization, and underlying event, with parameter values set according to the A14 parton-shower tune~\cite{ATL-PHYS-PUB-2014-021} and using the \NNPDF[2.3lo]~\cite{nnpdf:2012cx} set of PDFs.
The appropriate next-to-next-to-leading-order (NNLO) to LO $K$-factor of 1.3 is used to correct the MC LO signal cross-sections~\cite{Grazzini_2015,h2020resummed}.
Benchmark signal samples are generated in the mass range of 5--81~\GeV
following the previous 4$\mu$ search~\cite{HDBS-2018-57} and to ensure a negligible \Zp\ width compared with the detector resolution.
The contribution from $pp \to \Zp \tau \bar{\nu}_{\tau}$ to this search is found to be negligible and thus is not included in the MC signal samples.

The dominant SM background processes, \mbox{ $q\bar{q'} \rightarrow W(Z/\gamma^*)\rightarrow \ell^{\pm}\nu \ell^{\pm}\ell^{\mp}$ }($l=e, \mu,
\tau$; $\nu=\nu_e, \nu_\mu, \nu_\tau$; referred to as \qqlvll) and \mbox{$q\bar{q}\rightarrow (Z/\gamma^*)(Z/\gamma^*)\rightarrow \ell^{+}\ell^{-}\ell^{+}\ell^{-}$}(referred to as \qqllll), are
simulated with the \SHERPA[2.2.2] event generator~\cite{Bothmann:2019yzt}.
Matrix elements are calculated at next-to-leading-order~(NLO) accuracy in QCD for up to one additional parton and at LO accuracy for two and three additional parton emissions.
The ME calculations are matched and merged with the \SHERPA parton shower based on Catani–Seymour dipole factorization~\cite{Gleisberg:2008fv,Schumann:2007mg},
using the MEPS@NLO prescription~\cite{Hoeche:2011fd,Hoeche:2012yf,Catani:2001cc,Hoeche:2009rj}.
\SHERPA[2.2.2] is also used for the $gg\rightarrow \ell^{+}\ell^{-}\ell^{+}\ell^{-}$ process, with LO precision for zero- and one-jet final states, where a constant $K$-factor of 1.7~\cite{Caola_2015} is applied to account for NLO effects on the cross-section.
The events with non-prompt leptons arising from \Zjets\ and \ttbar\ processes are modeled using \SHERPA[2.2.1] and \POWHEGBOX[v2] generators~\cite{Frixione:2007nw,Nason:2004rx,Frixione:2007vw,Alioli:2010xd}, respectively.
Samples for other subdominant processes, such as the resonant $H\rightarrow ZZ^{*} \rightarrow 4\ell$ process, triboson processes ($VVV,V=W,Z$) and $W/Z$ bosons produced along with a \ttbar\ pair ($t\bar{t}V$), are also simulated as described in Ref.~\cite{HDBS-2018-57} and their contribution is found to be negligible and thus ignored.

Except for the signal, all samples are produced with a detailed simulation of the ATLAS detector~\cite{SOFT-2010-01} based on \GEANT~\cite{Agostinelli:2002hh}
to produce predictions that can be compared with the data.
The signal samples are processed through a faster simulation where the full Geant4 simulation of the calorimeter response is replaced by a parameterization of the shower shapes~\cite{SOFT-2010-01}.
Furthermore, simulated inelastic minimum-bias events are overlaid to model additional $pp$ collisions in the same and neighboring bunch crossings (\pileup)~\cite{SIMU-2020-01}.
Simulated events are reweighted to match the pileup conditions in the data. All simulated events were processed using the same reconstruction algorithms as used for data.


\section{Event reconstruction and selection}
%

%

Interaction vertices from the $pp$ collisions are reconstructed from at least two tracks with $\pT > 500$~MeV that are consistent with originating from the beam collision region in the $x$-$y$ plane. If more than one primary vertex candidate is found in the event, the candidate for which the associated tracks form the largest sum of squared \pT is selected as the hard-scatter primary vertex.

Muon candidates are reconstructed by combining the information from both the muon spectrometer (MS) and the inner detector (ID).
Muons are required to satisfy the `Medium' identification criterion~\cite{MUON-2018-03} and have $\pt > 3$~\GeV and $|\eta| < 2.5$.
Electrons and jets are used to select control samples for the background estimate.
Electrons are reconstructed from energy clusters in the electromagnetic calorimeter matched to ID tracks.
Candidate electrons must satisfy the `Tight' likelihood identification criterion~\cite{EGAM-2018-01} and have $\pt > 4.5$~\GeV and $|\eta| < 2.47$, excluding the transition region between the barrel and endcaps in the calorimeter ($1.37<|\eta|<1.52$).
All muons and electrons must be isolated from other particles based on a particle-flow algorithm
and satisfy the `PflowLoose' and `PflowTight' isolation criteria
~\cite{EGAM-2018-01,MUON-2018-03}, respectively.
Furthermore, muons~(electrons) are required to have matched tracks satisfying $|d_{0}|/\sigma_{d_{0}} < 3~(5)$ and $|z_{0} \sin(\theta)| < 0.5$~mm, where $d_{0}$ is the transverse impact parameter relative to the beam line, $\sigma_{d_{0}} $ is its uncertainty, and $z_{0}$ is the longitudinal impact parameter relative to the primary vertex.

Jet candidates are reconstructed from particle flow objects~\cite{PERF-2015-09} using the \antikt algorithm with a radius parameter of $R=0.4$~\cite{Cacciari:2008gp,Fastjet}.
They are calibrated using simulation with corrections obtained from in situ techniques in data~\cite{JETM-2018-05}.
A jet vertex tagger algorithm~\cite{PERF-2014-03} is applied to suppress pileup jets. All jets must have $\pT>30$~\GeV and $|\eta|<4.5$.
Jets containing $b$-hadrons, referred to as \bjets, are identified using a deep-learning neural network, \texttt{DL1r}~\cite{FTAG-2019-07}.
The chosen working point has an efficiency of 85\% for selecting \bjets with \pT $> 20$~\GeV and $|\eta| < 2.5$ and a rejection factor of about 3 and 40 for charm-jets and light-flavor jets, respectively~\cite{FTAG-2019-07}.

The missing transverse momentum, $\vec{p}_{\text{T}}^{\text{miss}}$~(with magnitude \MET), is defined as the negative vector sum of the \pT\ of all selected and calibrated objects in the event, including a term to account for the momentum from soft particles in the event that are not associated with any of the selected objects~\cite{PERF-2016-07}.

Events are required to contain exactly three muon candidates satisfying the selection criteria previously described.
Events with a fourth muon candidate satisfying a looser selection criterion as defined in the $4\mu$ channel~\cite{HDBS-2018-57}
are rejected to ensure the two search regions are disjoint.
Candidate events are required to have a total charge from the muons equal to $\pm1$. The three \pT-ordered muons are required to satisfy \pT thresholds of 20, 10, and 7~\GeV, respectively. The muons firing triggers are also required to satisfy the corresponding trigger \pT\ thresholds.
Events are required to have $\MET > 15$~\GeV. To suppress the background contribution from top-quark production, events containing $b$-jets are rejected.
Two opposite-sign muon pairs ($\mu^+\mu^-$) are selected from the three muons in each event.
Both the $\mu^+\mu^-$ pair must have an invariant mass greater than 4~\GeV to suppress the background contribution from low-mass resonances.
The $\mu^+\mu^-$ pair with the largest mass is referred as the leading pair $Z_1$, and the other pair is referred to as  the sub-leading pair $Z_2$.
Events are required to satisfy $m_{Z_1} < 85$~\GeV to suppress the background from $Z$-boson production.
These selection requirements define the signal region~(SR).


\section{Background estimation}
%

%
Background sources are classified into two categories:
the irreducible background with events containing prompt muons, and the reducible background with events
containing at least one non-prompt muon from hadron decays or misidentification of jets.

The irreducible background, which mainly originates from diboson production of \mbox{ \qqlvll} ~and~\mbox{\qqllll}, is estimated by using simulation.
The contribution from $t\bar{t}V$, $VVV$ and Higgs boson production processes is found to be negligible.
The background event yield of the dominant contribution from \qqlvll~production is normalized to data with
the help of a control region (CR) enriched in $\ell^{\pm}\nu\ell^{\pm}\ell^{\mp}$ events,
referred to as \WZCR.
The \WZCR\ sample is defined by selecting events with three leptons, where the lowest-\pT lepton must be a muon, satisfying the \pT\ requirements of 25, 20, and 20~\GeV, respectively,
and with $\MET > 25$~\GeV.
The same selection of lepton pairs used for the SR is implemented in the \WZCR, and the leading lepton pair forming the $Z$-boson candidate must have an invariant mass in the range of 85--100~\GeV.

The reducible background, with contributions from \Zjets, $Z+\gamma$, and \ttbar production processes,
is estimated by using a fake-factor method as described in
Refs.~\cite{STDM-2016-15,EGAM-2019-01}.
The fake factor is defined as the ratio of numbers of non-prompt muons $N^\mathrm{tight}_\mathrm{fake}/N^\mathrm{loose}_\mathrm{fake}$,
where \enquote{tight} or \enquote{loose} indicate whether those muons satisfy the impact parameter and isolation requirements,
or fail to meet at least one of the requirements.
The fake factor is measured in \Zjet events, considering an additional muon candidate that does not originate from the $Z$-boson decay.
The measurement is performed in bins of \pT of the additional muon and \MET.
The non-prompt muon background is then estimated by applying the fake factor as a weight to events satisfying the same selection as the SR, but with at least one loose-not-tight muon required.
The modeling of the estimated reducible background is studied in a validation region~(VR), which is disjoint to both the SR and the \WZCR.
The VR is defined using the same selections used for the SR but with two opposite-sign electrons with $\pT > 20$~\GeV
and a muon that satisfies the \enquote{tight} identification criteria.
The non-prompt muon background in this VR is also estimated with the fake-factor method.
The sum of the estimated non-prompt background yield and the MC prediction is consistent with data within the statistical uncertainties.


\section{Event classification with a parameterized deep neural network}
%

The signal and background have different distributions for the various kinematic variables.
A \textit{parameterized deep neural network}~(pDNN)~\cite{Baldi_2016}
is used to combine several discriminating variables into a single final discriminant.
The pDNN architecture allows the training of a single classifier for multiple signal mass hypotheses in the search range by adding a mass parameter together with other inputs.
The mass parameter is equal to the value of the nominal generated \Zp mass for the signal component, while a random value is drawn from the same distribution for the background mass parameter.
In the evaluation process, when applying the training results to real data, the mass parameter takes the value of the investigated signal mass.
The algorithm was implemented in the PyTorch~\cite{pytorch} framework.
Two classifiers are trained for low (high) \Zp mass searches using mass parameters less than (greater than or equal to) 40~\GeV.
A set of kinematic distributions was used for pDNN training input features:
the \pT of each muon, the invariant mass of the $Z_1$, $Z_2$ and three-muon system, $\Delta\phi$ of each muon pair that forms the $Z_1$ and $Z_2$,
\MET, \HT, which is the scalar sum of \MET and the \pT of all muons, \VT~$=\sqrt{\HTx^{2}+\HTy^{2}}$, and \MT~$=\sqrt{\HT^{2}-\VT^{2}}$,
where \HTx(\HTy) is the scalar sum of \MET and the \pT of all muons along the direction of the $x(y)$-axis.
%


\section{Systematic uncertainties}
%

%
Systematic uncertainties due to imperfect modeling of the detector in the simulation or the underlying physics of each process
are also considered for the prediction of signal and background processes.

Experimental uncertainties originate mainly from \MET resolution and scale, measurements of muon momentum resolutions and scales, muon reconstruction and identification efficiencies, jet energy scale and resolution, and $b$-tagging efficiency.
Uncertainties due to the trigger selection efficiency and pileup correction are also considered.
In addition, the uncertainty in the combined 2015--2018 integrated luminosity is 0.83\%~\cite{DAPR-2021-01}, obtained using the LUCID-2 detector~\cite{Avoni_2018} for the primary luminosity measurements.
Overall, the total experimental uncertainty in the predicted yields is $5\%~(7\%$) for the signal (background with prompt muons).

The theoretical uncertainties in the signal, and the major prompt background due to the diboson processes, include the uncertainties in PDFs, QCD scales, and \alphas.
The PDF uncertainty is estimated following the PDF4LHC~\cite{Butterworth:2015oua} procedure.
The \alphas uncertainty is estimated by varying the nominal \alphas = 0.118 by its uncertainty of $\pm0.001$.
The QCD scale uncertainty is estimated by varying the renormalization and factorization scales, following the procedure described in Ref.~\cite{STDM-2017-03}.
The parton showering uncertainty is estimated by comparing events with different parton shower parameters in the \SHERPA\ MC samples.
The total theoretical uncertainties in the reconstructed event yields for the signal and the \qqlvll\ background processes are estimated to be 15\% and 13\%, respectively, dominated by the QCD scale uncertainty.
The interference effect between the signal and SM DY background is about 2\%, and this effect is accounted as an uncertainty affecting the signal.

Systematic uncertainties assigned to the reducible background, about 11\% in total, mainly account for the measurement of the fake factors,
the differences between the composition of the events with fake leptons between \Zjet\ events and the events in the SR,
and data statistical uncertainties in the dedicated region where fake factors are applied.
The overall impact of systematic uncertainties in the search sensitivity is a degradation on the expected cross-section limits up to 14\% in the mass range considered.


%
\section{Results}
%

%
%
A simultaneous profile binned maximum-likelihood fit~\cite{Moneta:2010pm,Verkerke:2003ir,Cranmer:1456844} to the distribution of pDNN score in the SR and the event yield of the \WZCR\ is performed to constrain uncertainties and obtain information about a possible signal.
The normalizations of both the signal and the \qqlvll\ background are allowed to vary freely in the fit.
The systematic uncertainties are modeled as nuisance parameters subject to Gaussian constraints in the likelihood fit.
The expected signal yields and pDNN scores are interpolated across the MC generated signal samples and are used in the fitting process.
The fit is independently performed for each signal mass point since the pDNN score depends on the value of \mZp under test.
The binning of the pDNN distribution varies with each signal mass point to limit the size of the MC statistical uncertainties to at most 20\% per bin and also to maximize the expected signal sensitivity.
Table~\ref{tab:SR-yield} shows the expected background and observed event yields in the SR after the background-only fit.
The normalization factor of the \qqlvll\ background is determined to be $0.92 \pm 0.08$ and $0.91 \pm 0.09$ when the pDNN score is obtained with
$\mZp = 19$~\GeV and $\mZp = 60$~\GeV, respectively.
The corresponding distributions of the pDNN score are presented in Figure~\ref{fig:pDNN_postfit}.

\begin{table}[ht]
\centering
{
\caption{
Summary of observed and expected background yields in the SR after the likelihood fit under the background-only hypothesis.
The \enquote{Non-prompt} represents the contribution from non-prompt muons.
The uncertainty in the total background yield can be smaller than the quadrature sum of the contributions because of correlations resulting from the fit.
The expected signal yield obtained using the theoretical cross-section for a benchmark point ($m_{Z'}=19$~\GeV, $g_{Z'}=0.0085$) is also shown with its pre-fit uncertainty.
\label{tab:SR-yield}}
\small
\begin{tabular}{ c | c | ccc | c }
\hline\hline
Data               & Total background  & \qqlvll & $q\bar{q}/gg \to \ell^{+}\ell^{-}\ell^{+}\ell^{-}$ & Non-prompt & Signal \\
\hline
3089               &  $3080 \pm 54$      &  $1125 \pm 75$ &  $396 \pm 51$     &  $1559 \pm 86 $  & $36.5 \pm 0.5 $ \\  %
\hline\hline
\end{tabular}
}
\end{table}

\begin{figure}[!ht]
\centering
\subfloat[]{
\includegraphics[width=0.45\textwidth]{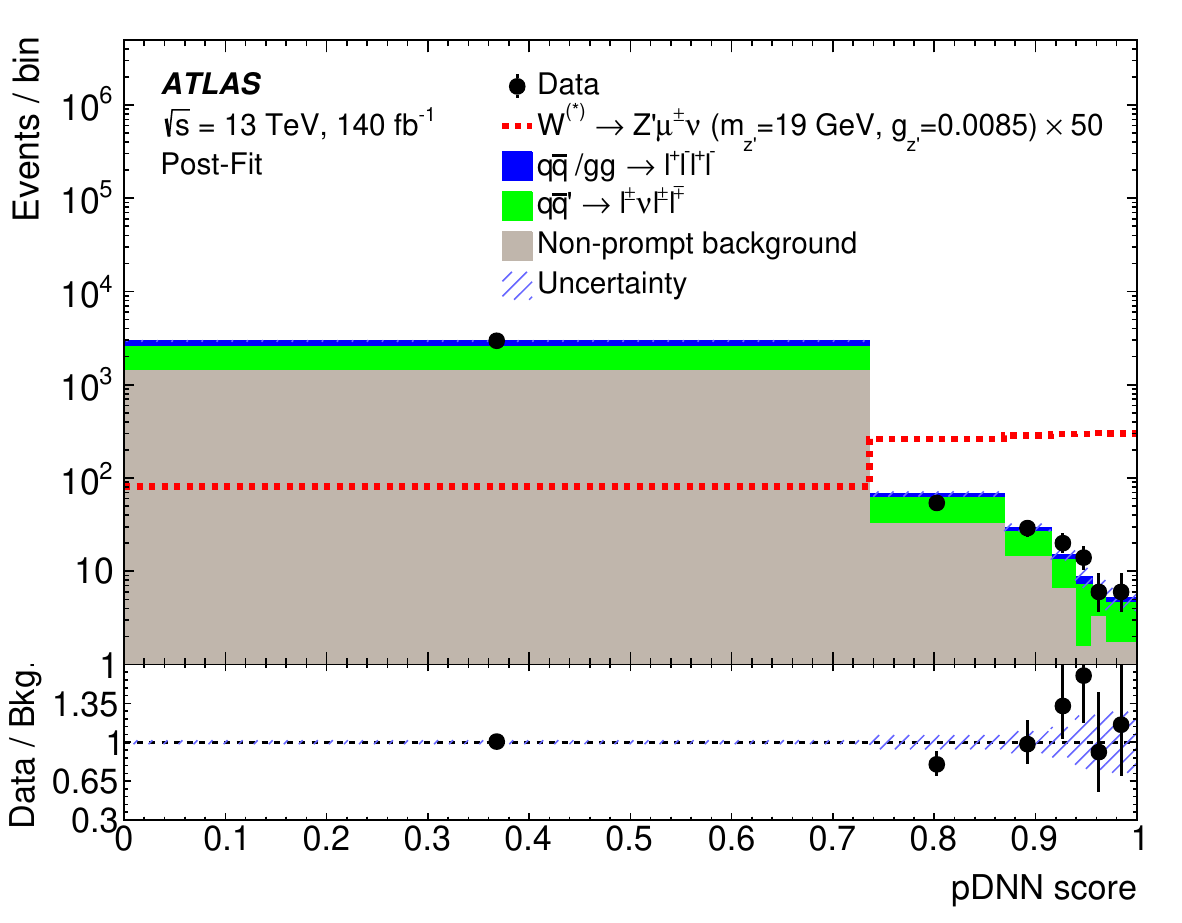}
\label{fig:pDNN_postfit:a}
}
\subfloat[]{
\includegraphics[width=0.45\textwidth]{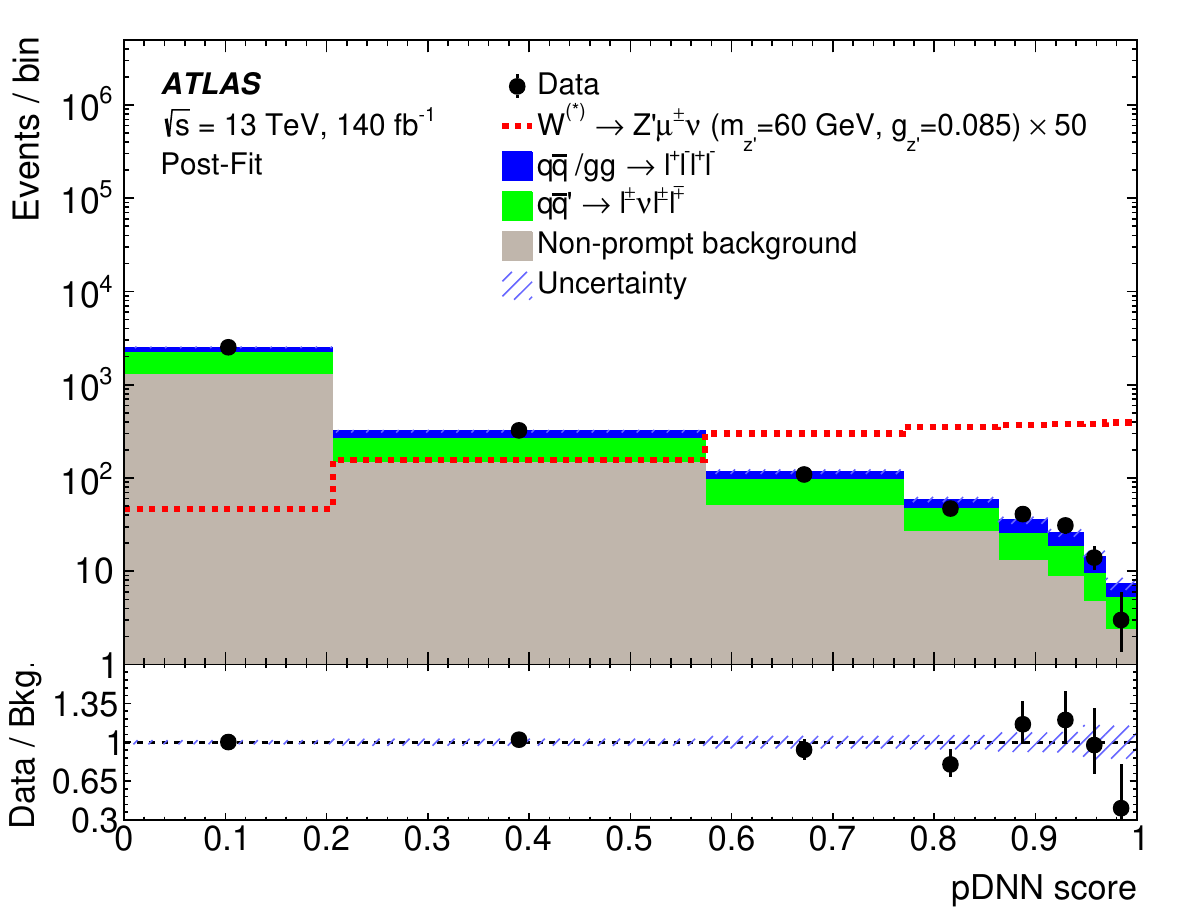}
\label{fig:pDNN_postfit:b}
}
\caption{The distributions of the pDNN score corresponding to (a) $\mZp = 19$~\GeV and (b) $\mZp = 60$~\GeV in the SR for the data and post-fit background contributions.
Signals are overlaid, with the predicted yield scaled up by a factor of 50.
The error bands include experimental and theoretical systematic uncertainties.
The ratio of the data to the background (\enquote{Bkg.}) prediction is shown in the lower panel.
}
\label{fig:pDNN_postfit}
\end{figure}

No significant deviation from the SM background hypothesis is observed and the largest excess of events is found for \mZp around 18.3~\GeV, with a local significance of $2.4\sigma$ and global significance of $0.8\sigma$.
Exclusion limits are set using the CL$_\mathrm{s}$ method~\cite{Read:2002hq}.
Upper limits at 95\% confidence level (CL) on the cross-section times branching fraction of the process \procWzp\ are shown in Figure~\ref{fig:limits}
as a function of \mZp.

\begin{figure}[!ht]
\centering
\includegraphics[width=0.43\textwidth]{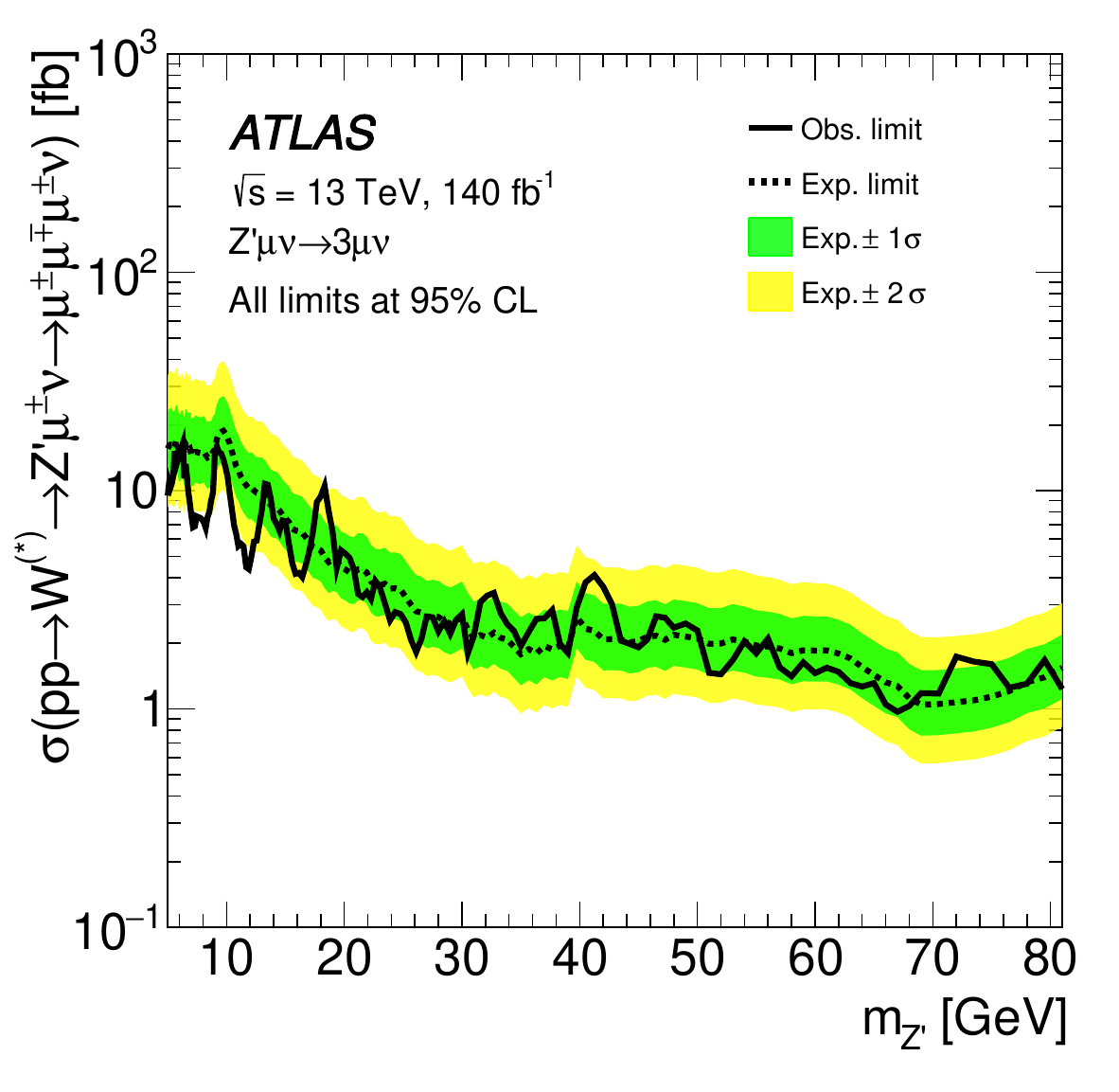}
\caption{Observed (solid line) and expected (dashed line) upper limits at 95\% CL on the production cross-section times branching fraction of the process \procWzp\ as a function of \mZp.
The surrounding shaded bands correspond to $\pm1$ and $\pm2$ standard deviations around the expected limit.
}
\label{fig:limits}
\end{figure}

A statistical combination with the ATLAS search for \Zp in the 4$\mu$ channel~\cite{HDBS-2018-57} is performed to improve the overall sensitivity.
The coupling parameter $g$ is used as a common parameter of interest for the $3\mu$ and $4\mu$ channels.
The contribution of the signal process \mbox{ \procZzp\ } in the SR of the $3\mu$ channel is also considered.
Common experimental uncertainties and theoretical modeling uncertainties are fully correlated.
The uncertainties relevant to backgrounds are uncorrelated due to a different background estimate in the $4\mu$ channel.
Upper limits at 95\% CL on the coupling parameter $g$ as a function of \mZp\ are shown in Figure~\ref{fig:limits_comb:a}.
The combined exclusion limits are significantly improved relative to the $4\mu$ channel. The improvement is up to 40\% in the high-mass region, where the $3\mu$ channel dominates the sensitivity.
In Figure~\ref{fig:limits_comb:b}, the results are also compared with the exclusion regions inferred from a measurement of neutrino tridents by the CCFR Collaboration~\cite{PhysRevLett.66.3117} and the $B_s$ mixing measurements by a global analysis performed in Ref.~\cite{Altmannshofer:2016jzy}.
The large region in the parameter space up to 81~\GeV allowed by the neutrino trident and $B_s$ measurements is now largely excluded.

\begin{figure}[!ht]
\centering
\subfloat[]{
\includegraphics[width=0.43\textwidth]{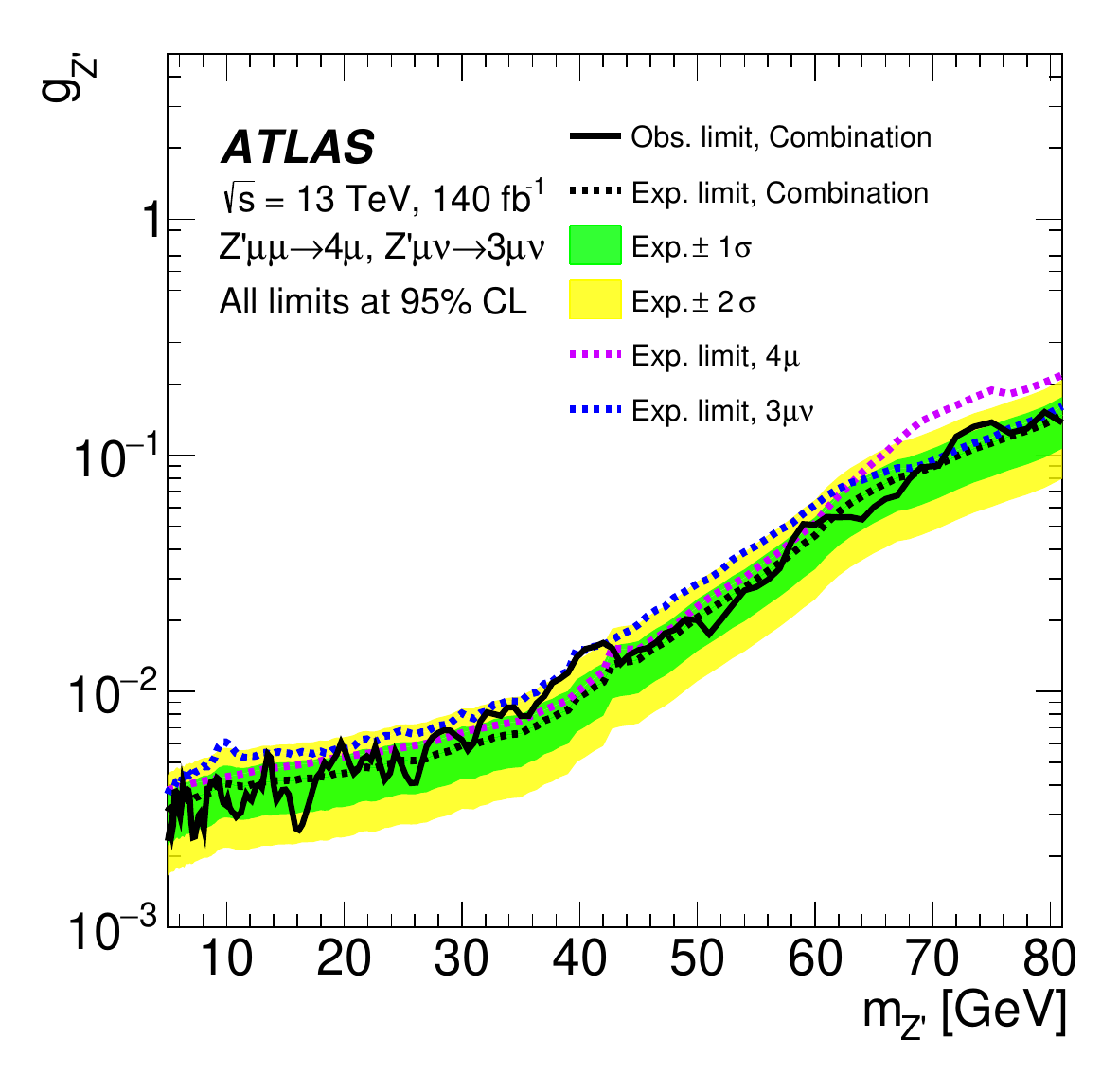}
\label{fig:limits_comb:a}
}
\subfloat[]{
\includegraphics[width=0.43\textwidth]{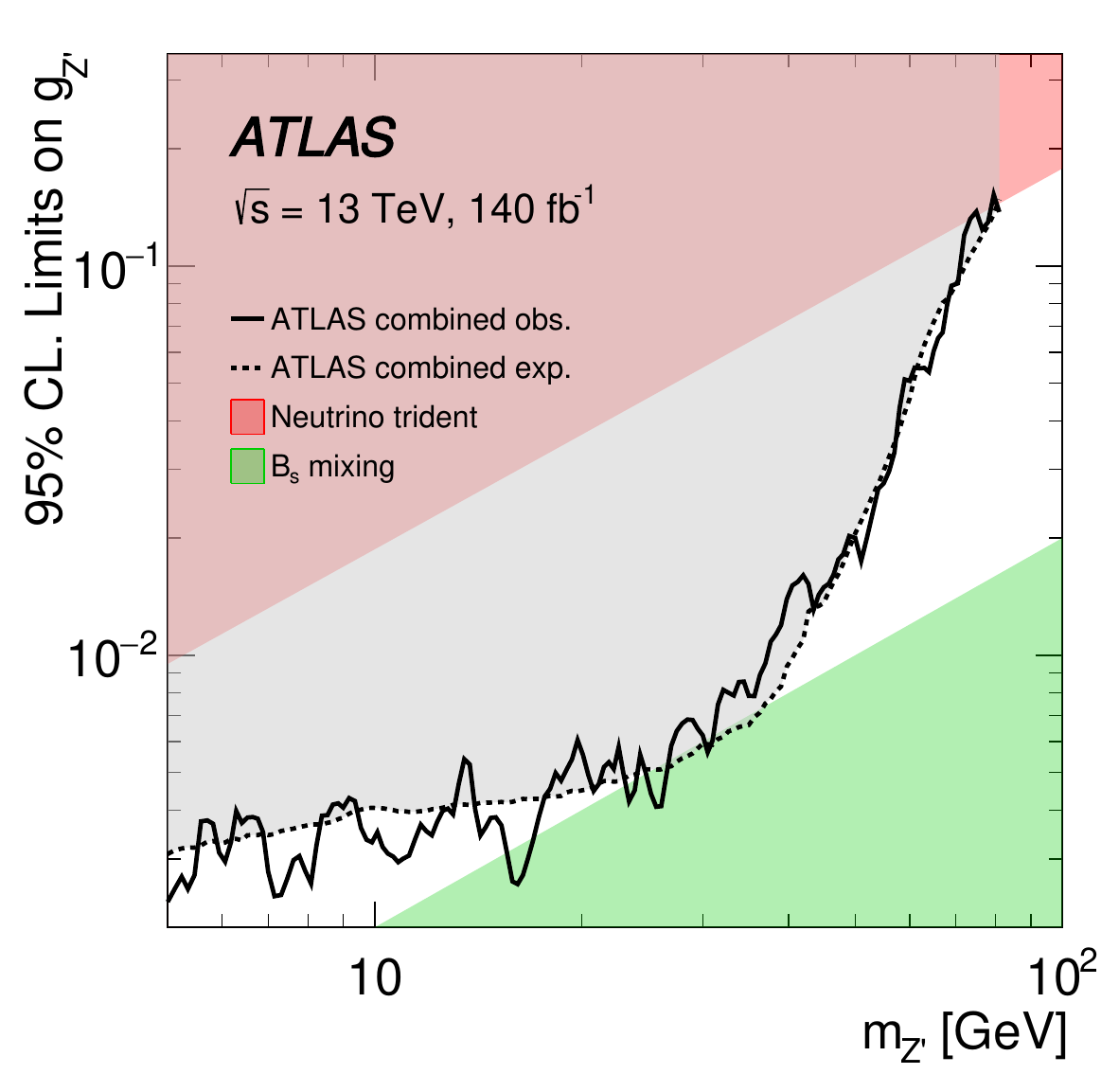}
\label{fig:limits_comb:b}
}
\caption{Results from the statistical combination of the $3\mu$ and $4\mu$ channels:
(a) observed and expected 95\% CL upper limits on $g$ as a function of \mZp,
(b) exclusion contour of $g$ compared with the limits inferred from the neutrino-trident (red) and the $B_s$ mixing (green) experimental results~\cite{Altmannshofer:2016jzy}.
}
\label{fig:limits_comb}
\end{figure}


%

%

%
%
\FloatBarrier

%
%
%

In conclusion,  the search for a \Lmt\ gauge boson \Zp using charged-current Drell--Yan production is reported for the first time at the LHC,
with a final state of $3\mu$ plus large missing transverse momentum,
using 140~\ifb{} of $\sqrt{s} = $ 13~\TeV{} proton--proton collision data collected with the ATLAS detector.
This search benefits from the considerably higher \Zp production cross-section compared with previous searches using the neutral-current Drell--Yan production,
and has better sensitivity for \mZp$>60$ GeV.
No significant excess of events over the expected SM background is observed.
Upper limits are set on the \Zp production cross-section times the decay branching fraction of the \procWzp\ process in a \Zp mass range of 5--81~\GeV.
The search is further statistically combined with the \Zp\ search using neutral-current Drell--Yan production with a $4\mu$ final state~\cite{HDBS-2018-57}.
The most stringent exclusion limits to date are set in the allowed parameter space of the \Zp coupling strength and \mZp.
%
%


%

%

%
\section*{Acknowledgments}
%

%
%

%
%

We thank CERN for the very successful operation of the LHC and its injectors, as well as the support staff at
CERN and at our institutions worldwide without whom ATLAS could not be operated efficiently.

The crucial computing support from all WLCG partners is acknowledged gratefully, in particular from CERN, the ATLAS Tier-1 facilities at TRIUMF/SFU (Canada), NDGF (Denmark, Norway, Sweden), CC-IN2P3 (France), KIT/GridKA (Germany), INFN-CNAF (Italy), NL-T1 (Netherlands), PIC (Spain), RAL (UK) and BNL (USA), the Tier-2 facilities worldwide and large non-WLCG resource providers. Major contributors of computing resources are listed in Ref.~\cite{ATL-SOFT-PUB-2023-001}.

We gratefully acknowledge the support of ANPCyT, Argentina; YerPhI, Armenia; ARC, Australia; BMWFW and FWF, Austria; ANAS, Azerbaijan; CNPq and FAPESP, Brazil; NSERC, NRC and CFI, Canada; CERN; ANID, Chile; CAS, MOST and NSFC, China; Minciencias, Colombia; MEYS CR, Czech Republic; DNRF and DNSRC, Denmark; IN2P3-CNRS and CEA-DRF/IRFU, France; SRNSFG, Georgia; BMBF, HGF and MPG, Germany; GSRI, Greece; RGC and Hong Kong SAR, China; ISF and Benoziyo Center, Israel; INFN, Italy; MEXT and JSPS, Japan; CNRST, Morocco; NWO, Netherlands; RCN, Norway; MEiN, Poland; FCT, Portugal; MNE/IFA, Romania; MESTD, Serbia; MSSR, Slovakia; ARIS and MVZI, Slovenia; DSI/NRF, South Africa; MICINN, Spain; SRC and Wallenberg Foundation, Sweden; SERI, SNSF and Cantons of Bern and Geneva, Switzerland; NSTC, Taipei; TENMAK, T\"urkiye; STFC/UKRI, United Kingdom; DOE and NSF, United States of America.

Individual groups and members have received support from BCKDF, CANARIE, CRC and DRAC, Canada; CERN-CZ, PRIMUS 21/SCI/017 and UNCE SCI/013, Czech Republic; COST, ERC, ERDF, Horizon 2020, ICSC-NextGenerationEU and Marie Sk{\l}odowska-Curie Actions, European Union; Investissements d'Avenir Labex, Investissements d'Avenir Idex and ANR, France; DFG and AvH Foundation, Germany; Herakleitos, Thales and Aristeia programmes co-financed by EU-ESF and the Greek NSRF, Greece; BSF-NSF and MINERVA, Israel; Norwegian Financial Mechanism 2014-2021, Norway; NCN and NAWA, Poland; La Caixa Banking Foundation, CERCA Programme Generalitat de Catalunya and PROMETEO and GenT Programmes Generalitat Valenciana, Spain; G\"{o}ran Gustafssons Stiftelse, Sweden; The Royal Society and Leverhulme Trust, United Kingdom.

In addition, individual members wish to acknowledge support from CERN: European Organization for Nuclear Research (CERN PJAS); Chile: Agencia Nacional de Investigaci\'on y Desarrollo (FONDECYT 1190886, FONDECYT 1210400, FONDECYT 1230812, FONDECYT 1230987); China: National Natural Science Foundation of China (NSFC - 12175119, NSFC 12275265, NSFC-12075060); Czech Republic: PRIMUS Research Programme (PRIMUS/21/SCI/017); EU: H2020 European Research Council (ERC - 101002463); European Union: European Research Council (ERC - 948254, ERC 101089007), Horizon 2020 Framework Programme (MUCCA - CHIST-ERA-19-XAI-00), European Union, Future Artificial Intelligence Research (FAIR-NextGenerationEU PE00000013), Italian Center for High Performance Computing, Big Data and Quantum Computing (ICSC, NextGenerationEU); France: Agence Nationale de la Recherche (ANR-20-CE31-0013, ANR-21-CE31-0013, ANR-21-CE31-0022, ANR-22-EDIR-0002), Investissements d'Avenir Labex (ANR-11-LABX-0012); Germany: Baden-Württemberg Stiftung (BW Stiftung-Postdoc Eliteprogramme), Deutsche Forschungsgemeinschaft (DFG - 469666862, DFG - CR 312/5-2); Italy: Istituto Nazionale di Fisica Nucleare (ICSC, NextGenerationEU), Ministero dell'Università e della Ricerca (PRIN - 20223N7F8K - PNRR M4.C2.1.1); Japan: Japan Society for the Promotion of Science (JSPS KAKENHI JP21H05085, JSPS KAKENHI JP22H01227, JSPS KAKENHI JP22H04944, JSPS KAKENHI JP22KK0227); Netherlands: Netherlands Organisation for Scientific Research (NWO Veni 2020 - VI.Veni.202.179); Norway: Research Council of Norway (RCN-314472); Poland: Polish National Agency for Academic Exchange (PPN/PPO/2020/1/00002/U/00001), Polish National Science Centre (NCN 2021/42/E/ST2/00350, NCN OPUS nr 2022/47/B/ST2/03059, NCN UMO-2019/34/E/ST2/00393, UMO-2020/37/B/ST2/01043, UMO-2021/40/C/ST2/00187, UMO-2022/47/O/ST2/00148); Slovenia: Slovenian Research Agency (ARIS grant J1-3010); Spain: BBVA Foundation (LEO22-1-603), Generalitat Valenciana (Artemisa, FEDER, IDIFEDER/2018/048), La Caixa Banking Foundation (LCF/BQ/PI20/11760025), Ministry of Science and Innovation (MCIN \& NextGenEU PCI2022-135018-2, MICIN \& FEDER PID2021-125273NB, RYC2019-028510-I, RYC2020-030254-I, RYC2021-031273-I, RYC2022-038164-I), PROMETEO and GenT Programmes Generalitat Valenciana (CIDEGENT/2019/023, CIDEGENT/2019/027); Sweden: Swedish Research Council (VR 2018-00482, VR 2022-03845, VR 2022-04683, VR grant 2021-03651), Knut and Alice Wallenberg Foundation (KAW 2017.0100, KAW 2018.0157, KAW 2018.0458, KAW 2019.0447); Switzerland: Swiss National Science Foundation (SNSF - PCEFP2\_194658); United Kingdom: Leverhulme Trust (Leverhulme Trust RPG-2020-004); United States of America: U.S. Department of Energy (ECA DE-AC02-76SF00515), Neubauer Family Foundation.

%
%


%
\clearpage
\part*{Appendix}

The \Zp mass resonance can be reconstructed either from $m_{Z_1}$ or $m_{Z_2}$ as shown in Figure~\ref{fig:app_1}, depending on the value of \mZp.

\begin{figure}[!htbp]
\centering
\subfloat[]{\includegraphics[width=0.43\textwidth]{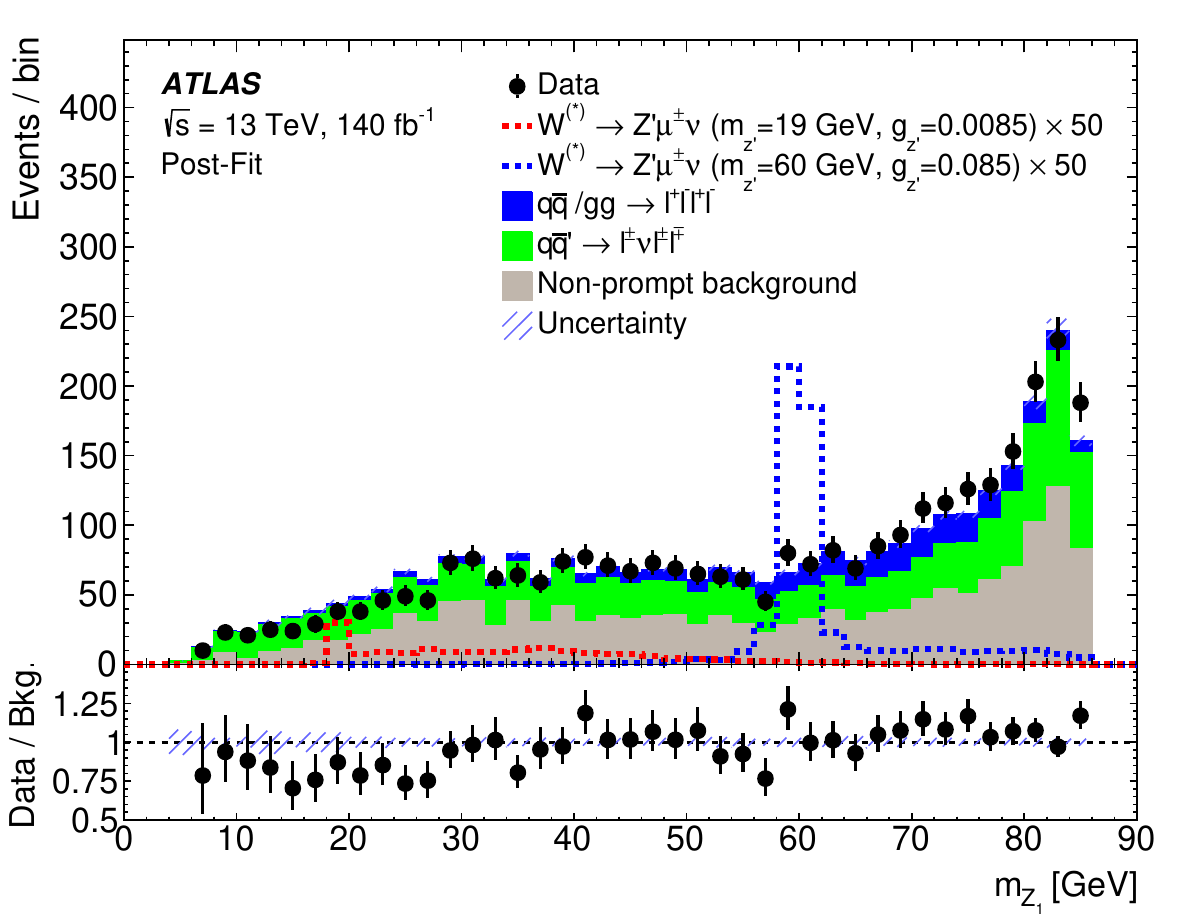}}
\subfloat[]{\includegraphics[width=0.43\textwidth]{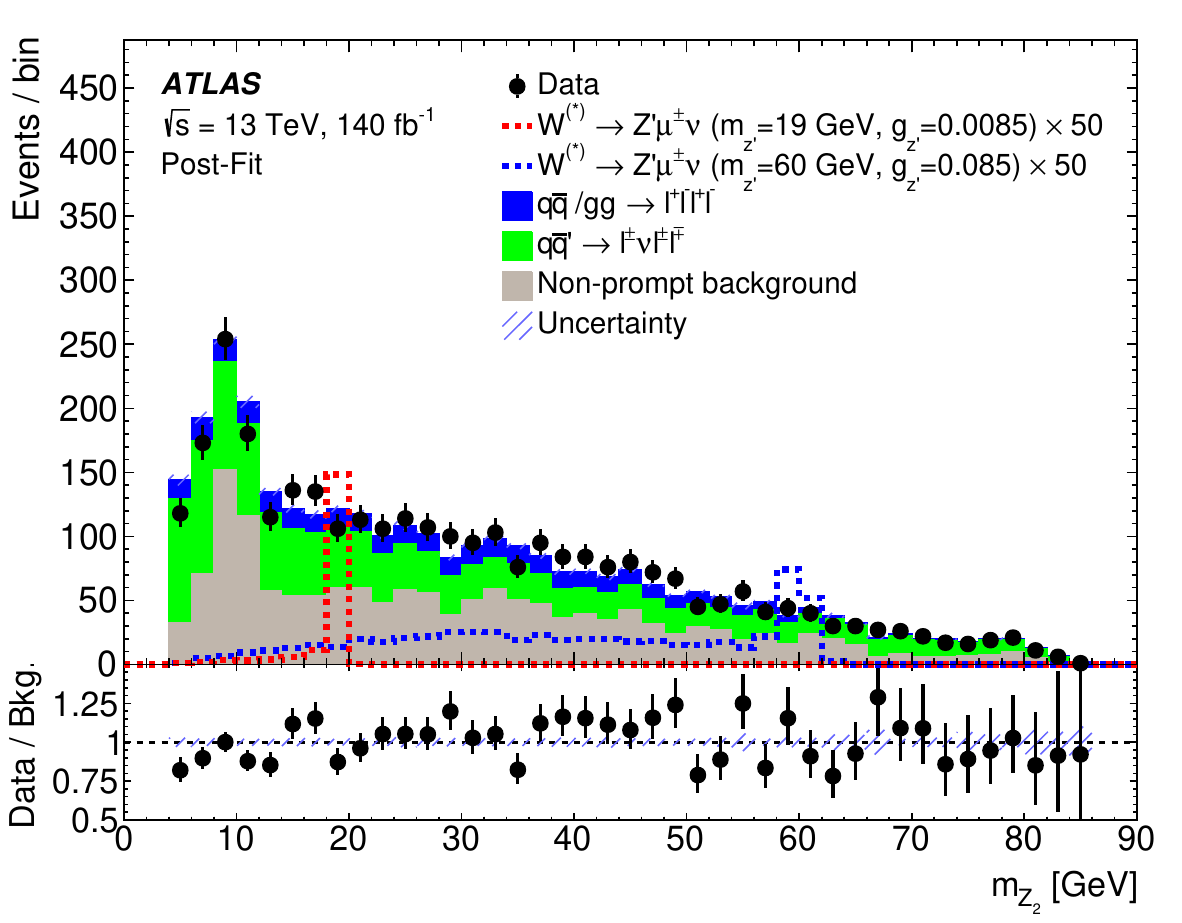}}
\caption{Comparison of data and post-fit SM prediction for: (a) the invariant mass of the leading $\mu^+\mu^-$ pair, $m_{Z_1}$, (b) the invariant mass of the sub-leading $\mu^+\mu^-$ pair, $m_{Z_2}$, in the SR\@. Two representative signals with masses of 19~\GeV and 60~\GeV are also overlaid, with the predicted yield scaled up by a factor of 50.
The error bands include experimental and theoretical systematic uncertainties.
The ratio of the data to the background (\enquote{Bkg.}) prediction is shown in the lower panel.
}
\label{fig:app_1}
\end{figure}


%
%

\clearpage
\printbibliography
\clearpage
 
\begin{flushleft}
\hypersetup{urlcolor=black}
{\Large The ATLAS Collaboration}

\bigskip

\AtlasOrcid[0000-0002-6665-4934]{G.~Aad}$^\textrm{\scriptsize 103}$,
\AtlasOrcid[0000-0001-7616-1554]{E.~Aakvaag}$^\textrm{\scriptsize 16}$,
\AtlasOrcid[0000-0002-5888-2734]{B.~Abbott}$^\textrm{\scriptsize 121}$,
\AtlasOrcid[0000-0002-1002-1652]{K.~Abeling}$^\textrm{\scriptsize 55}$,
\AtlasOrcid[0000-0001-5763-2760]{N.J.~Abicht}$^\textrm{\scriptsize 49}$,
\AtlasOrcid[0000-0002-8496-9294]{S.H.~Abidi}$^\textrm{\scriptsize 29}$,
\AtlasOrcid[0000-0002-9987-2292]{A.~Aboulhorma}$^\textrm{\scriptsize 35e}$,
\AtlasOrcid[0000-0001-5329-6640]{H.~Abramowicz}$^\textrm{\scriptsize 152}$,
\AtlasOrcid[0000-0002-1599-2896]{H.~Abreu}$^\textrm{\scriptsize 151}$,
\AtlasOrcid[0000-0003-0403-3697]{Y.~Abulaiti}$^\textrm{\scriptsize 118}$
% ...
\bigskip
\\

% ...
$^{16}$Department for Physics and Technology, University of Bergen, Bergen; Norway.\\
% ...
$^{29}$Physics Department, Brookhaven National Laboratory, Upton NY; United States of America.\\
% ...
$^{35}$$^{(a)}$Facult\'e des Sciences Ain Chock, Universit\'e Hassan II de Casablanca;$^{(b)}$Facult\'{e} des Sciences, Universit\'{e} Ibn-Tofail, K\'{e}nitra;$^{(c)}$Facult\'e des Sciences Semlalia, Universit\'e Cadi Ayyad, LPHEA-Marrakech;$^{(d)}$LPMR, Facult\'e des Sciences, Universit\'e Mohamed Premier, Oujda;$^{(e)}$Facult\'e des sciences, Universit\'e Mohammed V, Rabat;$^{(f)}$Institute of Applied Physics, Mohammed VI Polytechnic University, Ben Guerir; Morocco.\\
% ...
$^{49}$Fakult\"{a}t Physik , Technische Universit{\"a}t Dortmund, Dortmund; Germany.\\
% ...
$^{55}$II. Physikalisches Institut, Georg-August-Universit\"{a}t G\"ottingen, G\"ottingen; Germany.\\
% ...
$^{103}$CPPM, Aix-Marseille Universit\'e, CNRS/IN2P3, Marseille; France.\\
% ...
$^{118}$Department of Physics, New York University, New York NY; United States of America.\\
% ...
$^{121}$Homer L. Dodge Department of Physics and Astronomy, University of Oklahoma, Norman OK; United States of America.\\
% ...
$^{151}$Department of Physics, Technion, Israel Institute of Technology, Haifa; Israel.\\
$^{152}$Raymond and Beverly Sackler School of Physics and Astronomy, Tel Aviv University, Tel Aviv; Israel.\\
% ...
$^{a}$ Also Affiliated with an institute covered by a cooperation agreement with CERN.\\
$^{b}$ Also at An-Najah National University, Nablus; Palestine.\\
$^{c}$ Also at Borough of Manhattan Community College, City University of New York, New York NY; United States of America.\\
$^{d}$ Also at Center for High Energy Physics, Peking University; China.\\
$^{e}$ Also at Center for Interdisciplinary Research and Innovation (CIRI-AUTH), Thessaloniki; Greece.\\
$^{f}$ Also at Centro Studi e Ricerche Enrico Fermi; Italy.\\
% ...

\end{flushleft}


%

%
%

%

%
%
%
%

\end{document}